\documentclass[12pt]{iopart}
\usepackage{graphicx}
\begin{document}

\title[Transport in graphene nanoribbon]
  {Electronic and spin transport properties of graphene nanoribbon mediated by metal adatom: A study by QUAMBO-NEGF approach}

\author{G. P. Zhang$^{1,2}$, Xiaojie Liu$^{3,1}$, C. Z. Wang$^{1}$, Y. X. Yao$^{1}$, Jian Zhang$^{4}$ and K. M. Ho$^{1}$}
\address{$^{1}$Ames Laboratory, U. S. Department of Energy, and Department of Physics and Astronomy, Iowa State University, Ames, IA 50011, United States}
\address{$^{2}$Department of Physics, Renmin University of China, Beijing 100872, P. R. China}
\address{$^{3}$Beijing Computational Science Research Center, Beijing, 100084, P. R. China}
\address{$^{4}$3M Company, 3M Corporate Headquarters, 3M Center, St. Paul, MN 55144-1000, United States}
\ead{wangcz@ameslab.gov(C. Z. Wang);xiaojie@csrc.ac.cn(X. J. Liu)}

%%%%%%%%%%%%%%%%%%%%%%%%%%%%%%%%%%%%%%%%%%%%%%%%%%%%%%%%%%%%%%%%%%%%%
%% The document title should be given as usual. Some journals require
%% a running title from the author: this should be supplied as an
%% optional argument to \title.
%%%%%%%%%%%%%%%%%%%%%%%%%%%%%%%%%%%%%%%%%%%%%%%%%%%%%%%%%%%%%%%%%%%%%

%%%%%%%%%%%%%%%%%%%%%%%%%%%%%%%%%%%%%%%%%%%%%%%%%%%%%%%%%%%%%%%%%%%%%
%% Some journals require a list of abbreviations or keywords to be
%% supplied. These should be set up here, and will be printed after
%% the title and author information, if needed.
%%%%%%%%%%%%%%%%%%%%%%%%%%%%%%%%%%%%%%%%%%%%%%%%%%%%%%%%%%%%%%%%%%%%%
%\abbreviations{IR,NMR,UV}
%\keywords{graphene, metal adatom, QUAMBOs, electronic and spin transport}

%\begin{document}
%%%%%%%%%%%%%%%%%%%%%%%%%%%%%%%%%%%%%%%%%%%%%%%%%%%%%%%%%%%%%%%%%%%%%
%% The manuscript does not need to include \maketitle, which is
%% executed automatically.  The document should begin with an
%% abstract, if appropriate.  If one is given and should not be, the
%% contents will be gobbled.
%%%%%%%%%%%%%%%%%%%%%%%%%%%%%%%%%%%%%%%%%%%%%%%%%%%%%%%%%%%%%%%%%%%%%
\begin{abstract}
Structural and electronic properties, including deformation, magnetic moment, Mulliken population, bond order as well as electronic transport properties, of zigzag graphene nanoribbon (ZGNR) with Co adatom on hollow site are investigated by quasi-atomic minimal basis orbits (QUAMBOs), a first-principles tight binding (TB) scheme based on density functional theory (DFT), combined with non-equilibrium Green's function. As for electronic transport, below Fermi level the transmission is strongly suppressed and spin-dependent as a result of magnetism by Co adatom adsorption, while above Fermi level the transmission is slightly distorted and spin-independent. Due to local environment dependence of QUAMBOs-TB parameters, we construct QUAMBOs-TB parameters of ZGNR lead and ZGNR with Co adatom on the hollow center site by divide-and-conquer approach, and well reproduce the electronic transmission behavior. Our QUAMBO-NEGF method is a new and promising way for electronic transport in large-scale system.
\end{abstract}

\section{Introduction}
Graphene \cite{a1}, a true two-dimensional crystalline lattice formed by carbon atoms, has attracted tremendous interest in physics, chemistry, and material science. Its novel electronic structure, i.e., a linear dispersion near the Dirac point, generates unusual transport properties \cite{a1,a2,a3,a4,a5,a6,a7}, such as the conductivity linear with the gate voltage \cite{a1,a2}, unconventional integer quantum Hall effect \cite{a3,a4} and high carrier mobility \cite{a5}. In addition, the electronic, spintronics, transport, as well as magnetic properties of graphene can be modified through adsorption of various atoms and molecules on graphene interacting with graphene sheet \cite{a12,a13,a14,a15,a16,a17,a18}, therefore graphene can be applied as biosensor to gas molecular \cite{a8} and gate-tunable spin transport devices at room temperature \cite{a9,a10,a11}. The effects of graphite and metal contact on the transport properties are also important to future application of graphene-based materials \cite{PRL2008,ZhouMY-2010,transport-EPJB2006,transport-PRL2006,effectivemodel,ZhangGP-PLA2010,
ZhangGP-PLA2011,ZhangGP-CPL2011,ZhangGP-JPCM2012,Husj}. More interestingly, the strain can induce pseudo-magnetic field in graphene by recent studies \cite{YangHT-JPCM2011}.

First principle calculations based on density functional theory (DFT) had been widely proven to be very accurate for studying electronic properties of carbon-based system, such as carbon nanotube \cite{nanotube} and graphene. However, first principle calculations suffer from finite size effect due to huge computational demand for large-scale system. Therefore, lot of efforts had been made to extract tight binding (TB) parameters from first principle calculation \cite{ZhouMY-2010,Meyer-2011} to investigate electronic properties including electronic transport. Recently quasi-atomic minimal basis set orbitals (QUAMBOs) was developed by Yao et al., and TB Hamiltonian and overlap matrices of a given system can be accurately extracted from ab initio calculations without any fitting procedures \cite{a19,a20}. The Hamiltonian and overlap matrices in the QUAMBOs representation reproduce exactly the same energy bands up to a few eV above the Fermi level and wave functions of the occupied electronic states as those obtained by the fully converged first-principles calculations using the VASP \cite{a14,a19,a20}. Furthermore, the dependence on local environment of QUAMBOs-TB parameters enables the application of divide-and-conquer approach to study large scale system \cite{a19,carbonchain}. Finally, QUAMBO-NEGF scheme, i.e., QUAMBO approach combined with non-equilibrium Green's function (NEGF) method and Landauer formalism \cite{a21,a22,a23}, is applied to study the electronic transport in large atomistic systems \cite{carbonchain,a24}.

In this Letter, we first discuss structural and electronic properties, including deformation, magnetic moment, Mulliken population, as well as bond order, in the finite ZGNR with Co on hollow center (shortened as Co/ZGNR) and then investigate electronic transport through Co/ZGNR between ZGNR leads. This paper is arranged as follows. In Sec. 2, we present the VASP  (Vienna Ab initio Simulation Package) calculation details and formula of electronic transport based on QUAMBOs-TB Hamiltonian and overlap matrices. In Sec. 3, we discuss overall lattice distortion, magnetism, charge transfer and transport for Co/ZGNR. As for electronic transport, it is found that below Fermi level, the transmission is strongly suppressed and spin-dependent as a result of magnetism in system, while above Fermi level the transmission is slightly affected and almost spin-independent. Then we reconstruct QUAMBOs-TB parameters of ZGNR lead and Co/ZGNR by divide-and-conquer approach and the transmissions are well reproduced. Here we mainly present the method in this Letter, and detailed physical analysis of electron and spin transport in carbon-based system doped by metal adatom can be found in Refs. \cite{nanotube}. Finally we give a summary in Sec. 4.

\section{COMPUTATIONAL DETAILS}

The first-principles calculations are performed based on DFT with generalized gradient approximation (GGA) in the form of PBE (Perdew-Burke-Ernzerhof) \cite{a25} implemented in the VASP \cite{a26,a27,a28} code, including spin polarization and dipole moment corrections \cite{a29,a30}. Valence electrons are treated explicitly and their interactions with ionic cores are described by PAW (Projector Augmented Wave) pseudopotentials \cite{a31,a32}. The wave functions are expanded in a plane wave basis set with an energy cutoff of 600 eV. A k-point sampling of 6$\times$1$\times$1 Monkhorst-Pack grids in the first Brillouin zone of the supercell and a Gaussian smearing with a width of 0.05 eV is used in the calculations. All atoms in the supercell are allowed to relax until the forces on each atom are smaller than 0.01 eV/\AA. The supercell dimensions are kept fixed during the relaxation. For ZGNR lead, spin unpolarized basis is adopted in first-principles calculation \cite{carbonchain,a24}, since the ground state under spin polarized basis is not stable under finite temperature.

The Co/W$m$L$n$-ZGNR is modeled by having one Co adatom in a W$m$L$n$-ZGNR with periodic boundary condition along $x$-axis as shown in Fig. 1(a), in which $m$ and $n$ denote the number of carbon atoms along $x$- and $y$-axis respectively, and hydrogen atoms saturate the edges of ZGNR. For example, W4L8-ZGNR stands for ZGNR with $4$ and $8$ carbon atoms along $x$- and $y$-axis respectively. For graphene, the lattice constant obtained from our calculation is 2.46 \AA\ and agrees well with experimental value. The dimension of the supercell in the z direction is 15 \AA, which allows a vacuum region of about 12 \AA\ to separate the atoms in the supercell and their replicas. There are three typical sites for adatoms positioned near the center of ZGNR at the top of a carbon atom, labeled top (T) site, at the middle of a carbon-carbon bond, labeled bridge (B) site, and at the hexagonal center site, labeled hollow (H) site, respectively. In our calculations, H site is the energy stable site for Co adatom adsorbed as indicated in Fig. 1(a) on ZGNR.

Using the self-consistent wave functions from the plane-wave basis VASP calculations, a set of structure-dependent and highly localized QUAMBOs orbitals for Co/W$m$L$n$-ZGNR are constructed by following the scheme developed recently \cite{a19,a20}. Mulliken population analysis, bond order and QUAMBOs-TB parameters for such system are obtained by QUAMBOs scheme. In order to investigate the electronic transport properties of Co/W$m$L$n$-ZGNR, we define Co/W$m$L$n$-ZGNR as the conductor and set up a structure of lead-conductor-lead, as shown in Fig. 2(b) and Fig. 2(c). The lead is a semi-infinite W$m$-ZGNR and Fig. 2(a) shows one primitive unit (PU) of W4-ZGNR lead, which is determined by $R_{cutoff}$ and spacial localization of QUAMBOs \cite{a24}. Based on QUAMBOs-TB parameters, the electronic transmission with the spin index $\sigma$ in such Co/W$m$L$n$-ZGNR can be calculated as:

\begin{eqnarray}
T_{\sigma}(E)=Tr\left[\Gamma_{\textbf{L}}(E)G_{\sigma}^{R}(E)\Gamma_{\textbf{R}}(E)G_{\sigma}^{A}(E)\right],\\
\Gamma_{\textbf{L}}(E)=i\left[\Sigma_{\textbf{L}}^{R}(E)-\Sigma_{\textbf{L}}^{A}(E)\right].
\end{eqnarray}

Here $\Gamma_{\textbf{L}}(E)$ is the coupling function of the left ZGNR lead denoted as $\textbf{L}$, as shown in Fig. 2(b), and $\Sigma_{\textbf{L}}^{A}(E)$ is the conjugate of the advanced self energy $\Sigma_{\textbf{L}}^{R}(E)$. $\Sigma_{\textbf{L}}^{R}(E)$ is calculated through surface Green's function iteratively \cite{a21,a22,a23},
\begin{equation}
\Sigma_{\textbf{L}}^{R}(E)=\left(H_{\textbf{L}\textbf{M}}-ES_{\textbf{L}\textbf{M}}\right)g_{\textbf{L}}^{R}\left(H_{\textbf{L}\textbf{M}}^{\dag}-ES_{\textbf{L}\textbf{M}}^{\dag}\right),
\end{equation}
Here QUAMBOs-TB parameters at the lead-conductor interface, $H_{\textbf{L}\textbf{M}}$ and $S_{\textbf{L}\textbf{M}}$, should be spin-independent since enough buffer layers have been adopted in the calculation when Co adatom is adsorbed on ZGNR. $\textbf{M}$ stands for the conductor. $g_{\textbf{L}}^{R}$ is the surface Green's function of the left lead. $H_{\textbf{L}\textbf{M}}$ ($S_{\textbf{L}\textbf{M}}$) is the Hamiltonian (overlap) matrix between left lead and the conductor. The same calculation can also be performed for the right lead (denoted by $\textbf{R}$) to obtain $\Sigma_{\textbf{R}}^{R}(E)$. The retarded Green's function is calculated by
\begin{equation}
G_{\sigma}^{R}(E)=\left[ES_{\textbf{M},\sigma}-H_{\textbf{M},\sigma}-\Sigma_{\textbf{L}}^{R}(E)-\Sigma_{\textbf{R}}^{R}(E)\right]^{-1},
\end{equation}
where $H_{\textbf{M},\sigma}$ ($S_{\textbf{M},\sigma}$) is the Hamiltonian (overlap) matrix of the conductor, which should be spin-dependent for those atoms close to Co adatom and spin-independent for those atoms far away from Co adatom. $E$ is the energy for electron injecting from left reservoir and the energy reference (i.e., $E=0$) is chosen as the Fermi level of ZGNR lead, as the ZGNR lead is semi-infinite while Co/W$m$L$n$-ZGNR is finite. Since QUAMBOs-TB parameters near the lead-conductor interfaces are smooth when metal adatom has been screened, the difference between the Fermi level of ZGNR lead and that of Co/W$m$L$n$-ZGNR depends on the charge transfer and the interaction between Co adatom and W$m$L$n$-ZGNR. We stress that the conductance of system is under zero bias between left and right leads. Meanwhile, the density of states (DOS) $\rho_{\sigma}(E)=-Im Tr\left[G_{\sigma}^{R}(E)S_{\textbf{M},\sigma}\right]/\pi$ stands for Co/W$m$L$n$-ZGNR connected to ZGNR-lead, different from DOS in close Co/W$m$L$n$-ZGNR. Despite different calculation basis (i.e., spin-polarized vs spin-unpolarized) in ZGNR lead and different energy reference for electron transport, the trend of electron current with a certain dominant spin is consistent with the other similar work \cite{a16}.

\section{Results and discussion}
\subsection{Structural and electronic properties of Co/W$m$L$n$-ZGNR}
An important property induced by Co adatom adsorption on graphene nano-ribbon is the lattice deformation. In Fig. 1, the in-plane lattice distortion of ZGNR upon the adsorption of the Co adatom is plotted. In order to see the distortion more clearly, the displacements indicated by the blue arrows in the plots have been enlarged by a factor of 500 and 1,800 in Fig. 1 (b) and (c), respectively, for a better visualization. The distortion patterns (or strain fields) of the Co adatom adsorption on different types (W$m$L$n$) of ZGNR are similar as one can see from Fig. 1. The distortions of hydrogen atoms and carbon atoms at the edges of ZGNR are significant due to strong covalent bonds between hydrogen atoms and carbon atoms and the strain fields of each site both point to the Co adatom. The wider (in $y$ direction) of ZGNR, the larger of the distortion of carbon and hydrogen atoms is, which indicated that the strain fields of each site are repulsive to make the distortion diminished. We also examine the distortion of the carbon atoms nearby the Co adatom to see the type of strain fields. In Fig. 1(c), only the first and the second shell of carbon atoms away from Co adatom are shown. The amplitude of distortion of carbon atoms nearby the Co adatom is not as larger as those from the edge of ZGNR, where the displacements have been enlarged by 1,800. Note that the distortion pattern of Co adatom on ZGNR is different from Co on pure graphene \cite{a14}. The distortion patterns is emanative from the center for Co adatom on pure graphene, however, it is centralized for Co adatom on ZGNR.

The magnetic moment, charge transfer and the overall bond order of Co on various W$m$L$n$-ZGNR are listed in Table 1. We also calculate the magnetic moments for the Co adatom on ZGNR and compared them with those of the corresponding isolated atoms. There is about 1.0 $\mu_{B}$ of magnetic moment for Co adatoms on different types of ZGNR while the magnetic moment of an isolated Co atom is 3.0 $\mu_{B}$.  The origin of the magnetic moment reduction upon adsorption of Co adatom can also be attributed to the charge transfer. Our QUAMBOs \cite{a12,quambo-chargetranfer01,quambo-chargetranfer02} analysis shows that about 0.7$\sim$0.8 electrons transfer to ZGNR from Co adatom. We also examine the bond order between the Co adatom and carbon atoms in ZGNR to determine the bonding characters (i.e., ionic or covalent) in the systems. The sums of the bond orders of the whole system are presented in Table 1. Strong bond order values are obtained from our QUAMBOs analysis. The bond orders are as large as 1.7$\sim$2.0 for the overall interaction. The result indicates that the covalent interaction between the Co adatom and ZGNR is substantial. We note that the electronic properties of Co/W$m$L$n$-ZGNR have been converged as the length ($n$) and the width ($m$) of ZGNR increases, which one can see from Table 1. The results from our calculations suggest that the electronic wave functions and QUAMBOs-TB parameters obtained from the present calculations, which will be used in the quantitative transport analysis in the next section, are reliable. The local environment dependent QUAMBOs-TB parameters are guaranteed in such a post processing scheme.

\subsection{Electronic transport properties of ZGNR with Co adatom on hollow center}
For pure ZGNR lead, the transmission is determined by the number of eigenchannels available for a certain energy. During the sweep of the energy $E$, the transmission increases one when one eigenchannel turns on, and decreases one when one eigenchannel turns off. Interestingly, the Van Hove singularities of the density of states (DOS), as shown in Fig. 5(b), indicate turning on/off of an eigenchannel in W8-ZGNR. The transmission of W4-ZGNR, whose QUAMBOs-TB parameters are reconstructed as described below, is 1 within the energy range [-1.78eV, 1.92eV], except a very high transmission at Fermi level, as shown in Fig. 3. The high transmission of ZGNR was also found in the other similar work \cite{a33}. To investigate  electronic transport through Co/ZGNR, Co adatom should be screened by proper buffer layers between Co adatom and ZGNR lead, and the transmission through Co/W4L$n$-ZGNR is chosen to test the adequate buffer layers along $x$-direction. As the length of the conductor increases, i.e., $n$ varies from 8 to 16, the transmission curves collapses, as shown in Fig. 3(a) and Fig. 3(b). It indicates that $n=12$ is sufficient to include the required buffer layers and the interaction distance between Co adatom and carbon atoms is 7.2\AA. By the adsorption of Co adatom on the hollow center of ZGNR, the transmission is distorted and spin-dependent within some energy range, since a finite magnetic moment of 0.7$\sim$0.8 $\mu B$ is induced in the whole system. Within the energy range [-1.25eV, 0eV], the electronic transmission indexed by spin up dominates, while that indexed by spin down is much lower than 1. Within the energy range [0.5eV, 1.92eV], the spin-independent electronic transmission is perfect and 1, which is close to that through W4-ZGNR lead.

We discuss the tendency of electronic transport through different width of ZGNR with Co adatom on the hollow center, as shown in Fig. 3(c) and Fig. 3(d). Wider ZGNR usually provides more channels for electronic transport within the energy range [-2eV, 2eV], therefore the energy range corresponding to only one eigenchannel becomes narrower. As the width of the ZGNR-lead increases from 4 to 12, the above energy range is [-1.78eV, 1.92eV], [-1.06eV, 1.24eV] and [-0.76eV, 0.94eV] respectively. Similarly, the transmission at the Fermi level is larger than 1. The transmission through Co/W8L12- and Co/W12L12-ZGNR are similar to that through Co/W4L12-ZGNR. The transmission is strongly distorted by Co adatom adsorption and dominated by the component indexed by spin up below the Fermi level, while it is slightly distorted and spin-independent above the Fermi level.

\subsection{Electronic transport properties of ZGNR lead and Co/ZGNR with QUAMBOs-TB parameters constructed by divide-and-conquer approach}

The trend of electronic transport in wide ZGNR with Co adatom is highly desired, since electronic devices are usually wider than the system by first-principles calculations due to time and memory consuming. Here we demonstrate the application of divide-and-conquer approach to electronic transport of Co/ZGNR, based on local environment dependence of QUAMBOs-TB parameters, and the accuracy had been sufficiently proved in the application to the band structure of armchair-GNR \cite{a19} and to the electronic transmission of carbon atomic chain between ZGNR lead \cite{carbonchain}. First, we construct QUAMBOs-TB parameters of W$m$-ZGNR lead from the exact QUAMBOs-TB parameters of W8-ZGNR lead. As shown in Fig. 4(a), the atoms in ZGNR lead are classified as the edge atoms confined to the up and bottom frames and the center atoms denoted as $a$ and $b$, depending on their local environments. As demonstrated in Ref. \cite{carbonchain}, the $\pi$-orbit component of QUAMBOs-TB parameters of the carbon atoms $a$ and $b$ well represents those of all carbon atoms away from the edges. QUAMBOs-TB parameters of edge atoms in reconstructed W$m$-ZGNR lead are taken from the exact QUAMBOs-TB parameters of edge atoms in W8-ZGNR lead, while QUAMBOs-TB parameters of all other carbon atoms away from the edges in reconstructed W$m$-ZGNR lead are taken from those of carbon atoms $a$ and $b$ in W8-ZGNR lead.

We can get the electronic transmission as long as the QUAMBOs-TB parameters of reconstructed W$m$-ZGNR lead are obtained. For reconstructed W4-ZGNR lead, its transmission curve well consistent with that of the exact W4-ZGNR lead with $\pi$-orbit component (not shown here), since the electronic transport of pure ZGNR lead near the Fermi level are fully determined by delocalized $\pi$-orbit. For reconstructed W8-ZGNR lead, the transmission and density of states well reproduces that of exact W8-ZGNR lead, as shown in Fig. 5(a) and Fig. 5(b), except that within a narrow energy range [1.66eV, 1.8eV].  The transmission curve of Co/W8L12-ZGNR is also well reproduced, as shown in Fig. 5(c), when the exact W8-ZGNR lead is replaced by reconstructed W8-ZGNR lead. For reconstructed W12-ZGNR lead, the transmission is quantized and slightly asymmetric with respect to $E=0$. Compared with W8-ZGNR lead, since W12-ZGNR lead provides more channels for electronic transport, the energy range for each quantized transmission becomes narrower.

Now we further reconstruct QUAMBOs-TB parameters of Co/W12L12-ZGNR from those reconstructed parameters of atoms close to zigzag edges in W12L12-ZGNR and from those exact parameters of center atoms in Co/W12L12-ZGNR. There are two methods indicated by RCW8 and RCW10 to construct Co/W12L12-ZGNR, as shown in Fig. 4(c). The largest vertical distance between Co adatom and carbon atom is 8.4\AA\ and 10.5\AA\ respectively for RCW8 and RCW10 method. As Co adtom induces magnetism in ZGNR due to charge transfer, QUAMBOs-TB parameters of the center part in Co/W12L12-ZGNR system are spin-dependent, however spin splitting effect by Co adatom is slight for those carbon atoms close to the zigzag edges . Therefore it is reliable to adopt completely spin-independent QUAMBOs-TB parameters of atoms close to zigzag edges in W12L12-ZGNR to construct the corresponding part of Co/W12L12-ZGNR. The transmission curves of reconstructed Co/W12L12-ZGNR by RCW8 and RCW10 are quite accurate compared with the exact curve, as shown in Fig. 5(d). Similar to the interaction distance of Co adatom to ZGNR lead, i.e., 7.2\AA, RCW8 method is sufficient to reproduce the transmission.

Finally we note that the edges of actual ZGNR-samples are inevitably susceptible to various edge disorder, such as dangling bonds, vacancies and other possible defects, and the effect of edge disorder would certainly be interesting for further studying. QUAMBOs-TB parameters of ZGNR with edge disorder can be extracted from certain training cells and the accuracies of QUAMBOs-TB by divide-and-conquer approach can be assured as an example of armchair GNR was shown in Fig. 4 \cite{a19}. As for electron and spin transport, it should be affected by edge disorder as a result of scattering and the results will be different from those shown here.

\section{Summary}
Structural and electronic properties, including lattice deformation, magnetic moment, charge transfer, bond order and electronic transport properties of zigzag graphene nanoribbon (ZGNR) with Co adatom on hollow center are investigated. Due to a finite interaction distance between Co adatom and atoms in ZGNR embodied in magnetic moment and charge transfer, QUAMBOs-TB parameters are local environment dependent and are constructed by divide-and-conquer approach to investigate the electronic transmission. Our QUAMBO-NEGF method provides a promising way for study electronic transport in large-scale system.


%\begin{acknowledgement}
Acknowledgement:
Work at Ames Laboratory was supported by the US Department of Energy, Basic Energy Sciences, Division of Materials Science and Engineering including a grant of computer time at the National Energy Research Supercomputing Center (NERSC) in Berkeley, under Contract No. DE-AC02-07CH11358. Authors also acknowledged the support by the National Natural Science Foundation of China (Grant Nos.11174363, 11174366, 11204013 and 11204372).

%\end{acknowledgement}


\bibliography{QUAMBO-NEGF}

%% graphical table of contents.
%%%%%%%%%%%%%%%%%%%%%%%%%%%%%%%%%%%%%%%%%%%%%%%%%%%%%%%%%%%%%%%%%%%%%

\begin{figure}[tbh]
\begin{center}
\caption{(Color online) (a) The most stable structure of Co/W4L8-ZGNR. (b) The displacements have been enlarged by a factor of 500 in W$m$L$n$-ZGNR by Co adatom adsorption. (c) The displacements have been enlarged by a factor of 1800 of central atoms away from the Co adatom. The length of the arrows represents the amplitude of the distortion and the direction of the arrows represents the directions of the distortion. The displacements have been enlarged for a clear visualization.
}
\end{center}
\end{figure}

\begin{table} \caption{Magnetic moment ($\mu_{B}$), charge transfer ($e$), and bond order of Co on different types of graphene nano-ribbon (W$m$L$n$)}
\begin{tabular}{|c|c|c|c|}\hline
& Magnetic moment $\mu_{B}$ & Charge transfer ($e$) & Bond order \\ \hline
W4L8 & 1.11 & 0.70 & 2.04 \\ \hline
W4L12 & 1.06 & 0.81 & 1.90 \\ \hline
W4L16 & 1.04 & 0.81 & 1.89 \\ \hline
W8L12 & 1.10 & 0.75 & 1.76 \\ \hline
W12L12 & 1.13 & 0.70 & 1.65 \\ \hline
\end{tabular}
\end{table}

\begin{figure}[tbh]
\begin{center}
\includegraphics[width=6.0cm]{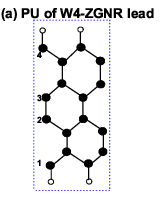}
\includegraphics[width=8.0cm]{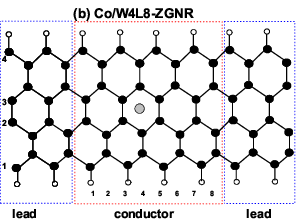}
\includegraphics[width=8.0cm]{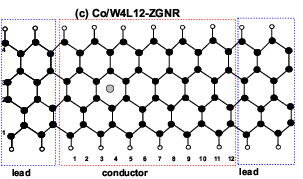}
\caption{(Color online) (a) One primitive unit (PU) of W4-ZGNR lead. Structure of lead-conductor-lead is set up to investigate the electronic transport. The lead is a semi-infinite W4-ZGNR, and the conductor is (b) Co/W4L8- and (c) Co/W4L12-ZGNR respectively. The black and white circles stand for carbon and hydrogen atoms respectively, and the grey circle on hollow center stands for Co adatom.}
\end{center}
\end{figure}

\begin{figure}[tbh]
\begin{center}
\includegraphics[width=8.0cm]{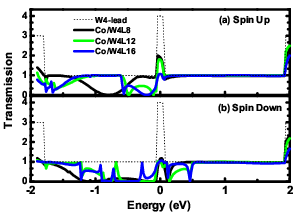}
\includegraphics[width=9.0cm]{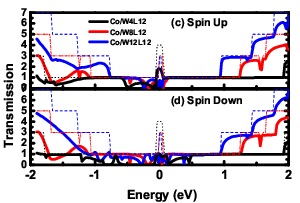}
\caption{(Color online) The electronic transmission indexed by (a) spin up and (b) spin down of W4-ZGNR lead, Co/W4L8-, Co/W4L12- and Co/W4L16-ZGNR. The electronic transmission indexed by (c) spin up and (d) spin down of Co/W4L12-, Co/W8L12- and Co/W12L12-ZGNR, and the transmissions of W4-, W8- and W12-ZGNR lead are also shown.}
\end{center}
\end{figure}

\begin{figure}[tbh]
\begin{center}
\includegraphics[width=8.0cm]{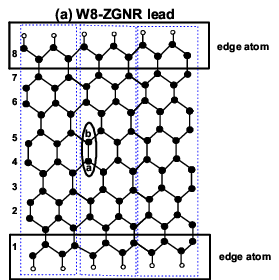}
\includegraphics[width=8.0cm]{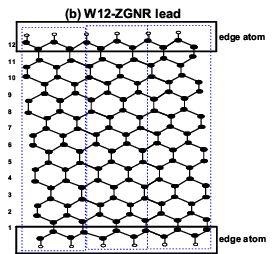}
\includegraphics[width=8.0cm]{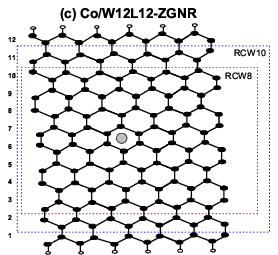}
\caption{(Color online) Illustration of divide-and-conquer approach to construct QUAMBOs-TB parameters in ZGNR lead and ZGNR with Co adatom on hollow center. (a) Three PUs of W8-ZGNR lead, (b) three PUs of W12-ZGNR lead, and (c) reconstruction of Co/W12L12-ZGNR. QUAMBOs-TB parameters of edge carbon atoms, edge hydrogen atoms and center carbon atoms ($a$ and $b$) are extracted from W8-ZGNR lead and then adopted to construct W8- and W12-ZGNR lead. RCW8 and RCW10 indicates that QUAMBOs-TB parameters of the center 8 and 10 lines of carbon atoms and Co adatom are extracted from Co/W12L12-ZGNR, while QUAMBOs-TB parameters of all other atoms are extracted from W12L12-ZGNR.}
\end{center}
\end{figure}

\begin{figure}[tbh]
\begin{center}
\includegraphics[width=10.0cm]{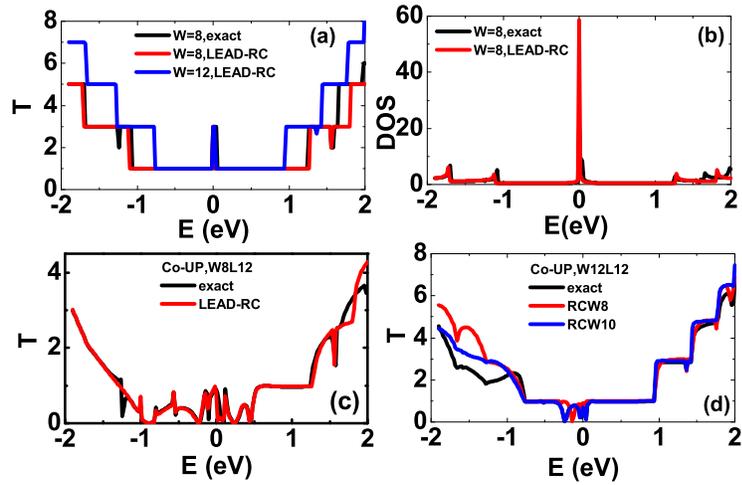}
\caption{(Color online) (a) The transmissions of the exact W8-ZGNR lead, the reconstructed W8- and W12-ZGNR lead. (b) The density of states of the exact and reconstructed W8-ZGNR lead. (c) The transmissions indexed by spin up of Co/W8L12-ZGNR between the exact and the reconstructed W8-ZGNR lead. (d) The comparison of the transmission indexed by spin up of Co/W12L12-ZGNR with the exact value by RCW8 and RCW10 respectively.}
\end{center}
\end{figure}

\end{document}